\documentclass[aps,prl,showpacs,twocolumn,floatfix]{revtex4}
\usepackage{amsmath,amsfonts,amssymb,graphicx,subfigure,float}

\newcommand{\qedsymbol}{$\blacksquare$}

\begin{document}

 \title{Decoherence of Fock States Leads to a Maximally Quantum State}
 \author{Andrew C. McClung}
 \author{Tyler E. Keating}
 \author{Adam T. C. Steege}
 \author{Arjendu K. Pattanayak}
 \affiliation{
 Department of Physics and Astronomy, Carleton College, 
 One North College Street, Northfield, MN 55057
 }
 \date{July 26, 2010}
 
 \begin{abstract}
  We consider the Wigner function evolution of Fock states $|n\rangle$ linearly coupled to a 
  Markovian bath of oscillators. In the absence of environmental coupling, apparent ``quantumness'' 
  increases with $n$, but the presence of {any} environmental interaction causes high-$n$ 
  states to lose their quantum features more rapidly than low-$n$ states. Using the negative volume 
  of the Wigner function as a metric \cite{kenfack04}, we observe a 
  time-dependent quantumness peak across the eigenstates.
 \end{abstract}
 
 
 \maketitle
 
 The macroscopic world is constructed of fundamentally quantum objects, but the observation of 
 quantum effects requires either very small or very cold systems. That is, quantum effects are most 
 easily visible in systems sufficiently isolated from their environments to prevent decoherence. 
 It is generally more difficult to adequately protect larger systems from decoherence; this difficulty 
 gives rise to the classical behavior of the macroscopic scale \cite{zurek03}. 
 Contemporary investigations of the transition from quantum to classical are motivated by
 fundamental considerations as well as practical issues of control and engineering.
 The transition is being investigated in a wide array of systems, including nanomechanical systems 
 \cite{blencowe04,brennecke08}, mesoscopic systems such as Josephson junction devices 
 \cite{wal00}, as well as cavity-QED systems \cite{mabuchi02}.
 
 This quantum-classical transition is understood to be a multi-parameter transition, controlled by 
 internal system parameters and behaviors as well as (a) the relative size of $\hbar$ compared to
 the characteristic action of the system, (b) the strength of the system-environment interaction, and
 (c) the temperature and other characteristics of the environment, leading to a rich landscape of 
 possible behaviors.
 
 In this paper we argue that the interplay between the relative effects of decoherence and 
 increasing quantum number leads to nonmonotonicity in the ``quantumness'' of a unipartite 
 system. We illustrate this in the behavior of Fock states $|n\rangle$ for a harmonic oscillator 
 coupled to a environment modeled as a Markovian bath of oscillators and demonstrate the 
 existence of a time-dependent quantumness peak. Signatures of this effect have been seen in 
 recent experimental work mapping the decoherence of Fock states \cite{brune08,wang08}, 
 although the significance had not been previously realized.
 
 In the interaction picture, the time evolution of the density matrix $\rho$ for a quantum harmonic 
 oscillator coupled to a continuum of oscillators in the Markovian 
 approximation and with appropriate assumptions about the spectrum of the environment is
 governed by the master equation \cite{gardiner00}
  \begin{equation}
   \dot{\rho} = \frac{\gamma}{2}\left(
    \bar{N}L[a^\dagger]\rho
    + (\bar{N}+1)L[a]\rho
   \right).
   \label{eq:masterEquation}
  \end{equation}
 Here the dot represents the time derivative, the Lindblad superoperator is defined as $L[O]\rho 
 \equiv2O\rho O^\dagger-O^\dagger O\rho-\rho O^\dagger O$, $a^\dagger$ and $a$ are the raising 
 and lowering operators for the harmonic oscillator, respectively, $\gamma$ represents the degree 
 of  coupling of the oscillator to its environment, and $\bar{N}$ corresponds to the mean number of 
 thermal photons in the bath.
 
 In our analysis below, we consider the Wigner function $W(\alpha)$, a phase space representation
 of the state of the system, which is defined most easily (if circuitously) via the Fourier transform of $
 \chi(\lambda)$, the Wigner characteristic function \cite{gardiner00}:
  \begin{equation}
   W(\alpha) = \iint \frac{d^2\lambda}{\pi}\chi(\lambda)\exp(\lambda^*\alpha - \lambda \alpha^*),
   \label{eq:wchi}
  \end{equation}
 where $\alpha=(x+ip)/\sqrt{2}$, in terms of the phase space quadratures $x$ and $p$; and 
 $\lambda=(\tilde x+i\tilde p)/\sqrt{2}$ is defined on the Fourier transform of phase space.
 The function $\chi(\lambda)$ is written in terms of the density matrix as $ \chi
 (\lambda) = \mathrm{Tr}\left[\rho \exp(\lambda a^\dagger -\lambda^*a)\right]$. The master equation 
 \eqref{eq:masterEquation} can then be rewritten \cite{serafini04} as a diffusion equation for the characteristic 
 function, $\chi$:
  \begin{equation}
   \dot{\chi} = -\frac{\gamma}{2}\left[
    \tilde x \frac{\partial \chi}{\partial \tilde x}
    +\tilde p \frac{\partial \chi}{\partial \tilde p}
    + \left(\frac{1}{2} + \bar{N}\right)(\tilde x^2 + \tilde p^2)\chi
   \right].
   \label{eq:diffEq}
  \end{equation}
 Restricting our attention to systems initialized in the $n$th Fock state, the solution to Eq. 
 \eqref{eq:diffEq} can be written \cite{serafini04}
  \begin{equation}
   \chi_n(\lambda, t) = L_n(|\lambda|^2e^{-\gamma t})
   e^{\frac{1}{2}|\lambda|^2(1-2\bar{N}(e^{-\gamma t}-1))},
   \label{eq:solns}
  \end{equation}
 where $L_n(x) = \sum_{m=0}^n\frac{(-x)^m}{m!}{n\choose m}$ is the Laguerre polynomial of order 
 $n$. 
 
Recall that in the absence of environmental interaction, 
the Wigner function for the $n$th Fock state is given by \cite{gardiner00}
  \begin{equation}
   \mathcal{W}_n(x,p) = \frac{(-1)^n}{\pi}e^{-(x^2+p^2)}L_n(2(x^2+p^2)).
   \label{eq:niwig}
  \end{equation}
Given Eq.\ \eqref{eq:niwig}, for zero temperature $(\bar{N} = 0)$, the time-dependent solution \eqref{eq:solns} for an 
environmentally-interacting harmonic oscillator becomes a time-dependent 
superposition of non-interacting Wigner functions $\mathcal{W}_m(x,p)$, where $m\leq n$:
  \begin{equation}
   W_n(x,p,t) = e^{-n\gamma t}\sum_{m=0}^n{n\choose m}(e^{\gamma t}-1)^{n-m}
   \mathcal{W}_m(x,p).
   \label{eq:iwig}
  \end{equation}
That is, the Wigner function of a decohering Fock state is composed exclusively of diagonal 
elements. 

To demonstrate Eq.\ \eqref{eq:iwig}, we note that substituting 
  Eq.\ \eqref{eq:solns} into Eq.\ \eqref{eq:wchi} yields
   \begin{equation}
    W_n(x,p,t) = \sum_{m=0}^n\frac{(-e^{-\gamma t})^m}{(2\pi)^2m!}{n\choose m}f_m(x,p),
    \label{eq:iwig2}
   \end{equation}
  where
   \begin{equation}
    f_m(x,p) = \iint d\tilde xd\tilde p|\lambda|^{2m}e^{\frac{1}{2}|\lambda|^2}\exp(\lambda^*\alpha - 
    \lambda \alpha^*).
    \label{eq:f}
   \end{equation} 
   
  \paragraph{Proposition.}We must now establish that
   \begin{equation}
    f_m(x,p) = (2\pi)^2m!\sum_{k=0}^{m}(-1)^k{m\choose k}\mathcal{W}_k(x,p).
    \label{eq:prop1}
   \end{equation}
   
  \paragraph{Proof.} We prove Eq. \eqref{eq:prop1} by induction. First, we rearrange 
  Eq. \eqref{eq:iwig2} to get a recursion relation $f_n(x,p)$, yielding,
   \begin{align}
    &f_n(x,p) = \frac{(2\pi)^2n!}{(-e^{-\gamma t})^n}\Bigg[
     W_n(x,p,t) \nonumber\\
     &\quad- \sum_{m=0}^{n-1}\frac{(-e^{-\gamma t})^m}{(2\pi)^2m!}{n\choose m}f_m(x,p)
    \Bigg]
    \label{eq:fn}
   \end{align}
  for $n>0$ and
   \begin{equation}
    f_0(x,p) = (2\pi)^2W_0(x,p,t),
    \label{eq:baseCase}
   \end{equation}
  which is the base case ($n=0$) of Eq. \eqref{eq:prop1}. To prove 
  that \eqref{eq:prop1} holds true in general, we now 
  show that Eq. \eqref{eq:prop1} holds for the $n+1$th case. We consider 
  $f_{n+1}(x,p)$ by evaluating Eq. \eqref{eq:fn} at $t=0$:
   \begin{align}
    &f_{n+1}(x,p) = \frac{(2\pi)^2(n+1)!}{(-1)^{n+1}}\Bigg[
     W_{n+1}(x,p,t) \nonumber\\
     &\quad- \sum_{m=0}^{n}\frac{(-1)^m}{(2\pi)^2m!}{n+1\choose m}f_m(x,p)
    \Bigg].
    \label{eq:fn+1}
   \end{align}
   Because the sum only considers $f_m(x,p)$ where $m\leq n$, we may employ Eq. 
   \eqref{eq:prop1}, giving
    \begin{align}
     &f_{n+1}(x,p) ={(2\pi)^2(n+1)!}\Bigg\{\Big[{(-1)^{n+1}}
     W_{n+1}(x,p,t)\Big] \nonumber\\
     &\quad+ \sum_{m=0}^{n}\sum_{k=0}^{m}{(-1)^{k+m+n}}{n+1\choose m}
     {m\choose k}\mathcal{W}_k(x,p)
    \Bigg\}.
    \label{eq:fn+12} 
    \end{align}
   The term in the square brackets is the $n+1$th term of the sum in Eq. \eqref{eq:prop1}; we 
   must demonstrate is that the double sum in Eq. \eqref{eq:fn+12} constitutes the first $n$ terms. 
   The double sum can be rewritten as
    \begin{align}
     &\sum_{m=0}^{n+1}\sum_{k=0}^{m}(-1)^{k+m+n}{n+1\choose m}{m\choose k}
     \mathcal{W}_k(x,p)\nonumber\\
     &\quad+\sum_{k=0}^{n}(-1)^{k}{n+1\choose k}
     \mathcal{W}_k(x,p)\nonumber\\
     &\quad-(-1)^{n}\mathcal{W}_{n+1}(x,p).
     \label{eq:dsrw}
    \end{align}
   The second term in Eq. \eqref{eq:dsrw} constitutes the first $n$ terms of the sum in Eq. 
   \eqref{eq:prop1} evaluated for $n+1$. All that remains to be shown to establish Eq. 
   \eqref{eq:prop1} is that the first and third terms in expression \eqref{eq:dsrw} cancel exactly;
   that is, we must show
    \begin{equation}
     \sum_{m=0}^{n+1}\sum_{k=0}^{m}(-1)^{k+m}{n+1\choose m}{m\choose k}
     \mathcal{W}_k(x,p) = \mathcal{W}_{n+1}(x,p).
     \label{eq:prop2}
    \end{equation}
   The LHS of Eq. \eqref{eq:prop2} can be rewritten as
    \begin{equation}
     \sum_{m=0}^{n+1}\mathcal{W}_m\sum_{k=m}^{n+1}(-1)^{k+m}{n+1\choose k}{k\choose m};
     \label{eq:dsrw2}
    \end{equation}
   but the second sum in Eq. \eqref{eq:dsrw2} is the well-known combinatorial identity
    \begin{equation}
     \sum_{k=a}^{b}(-1)^{k+a}{b\choose k}{k\choose a} = \delta_{a,b},
    \end{equation}
   where $\delta_{a,b}$ is the Kronecker delta \cite{benjamin03}; therefore the only surviving term in 
   the LHS of Eq. \eqref{eq:dsrw2} is $\mathcal{W}_{n+1}(x,p)$, thereby establishing Eq. 
   \eqref{eq:prop1}. \qedsymbol
     
   Inserting Eq. \eqref{eq:prop1} into Eq. \eqref{eq:iwig2} gives
    \begin{equation}
     W_n(x,p,t) = \sum_{m=0}^n\sum_{k=0}^{m}(-1)^{k+m}{e^{-m\gamma t}}{n\choose m}
     {m\choose k}\mathcal{W}_k(x,p);
     \label{eq:iwig3}
    \end{equation}
   by collecting like $\mathcal{W}_k(x,p)$, we can rewrite Eq. \eqref{eq:iwig3} as
    \begin{align}
     &W_n(x,p,t) = \sum_{m=0}^n\mathcal{W}_m(x,p){n\choose m}\nonumber\\
     &\quad\times\sum_{k=m}^{n}(-1)^{k+m}{e^{-k\gamma t}}
     \frac{(n-m)!}{(k-m)!(n-k)!}.
     \label{eq:iwig4}
    \end{align}
   By changing the index on the second sum to $j=k-m$, Eq. \eqref{eq:iwig4} becomes
    \begin{align}
     &W_n(x,p,t) = \sum_{m=0}^n\mathcal{W}_m(x,p){n\choose m}\nonumber\\
     &\quad\times e^{-n\gamma t}\sum_{j=0}^{n-m}(-1)^{j}{(e^{\gamma t})^{(n-m)-j}}
     {n-m\choose j};
     \label{eq:iwig5}
    \end{align}
   but the second sum in Eq. \eqref{eq:iwig5} is simply the expansion of $(e^{\gamma t}-1)^{n-m}$, 
   and hence Eq. \eqref{eq:iwig} is recovered.
  
  \begin{figure*}
   \begin{minipage}[b]{0.6\textwidth}
   \begin{flushleft}
   \subfigure[]{
    \includegraphics[width=\textwidth]{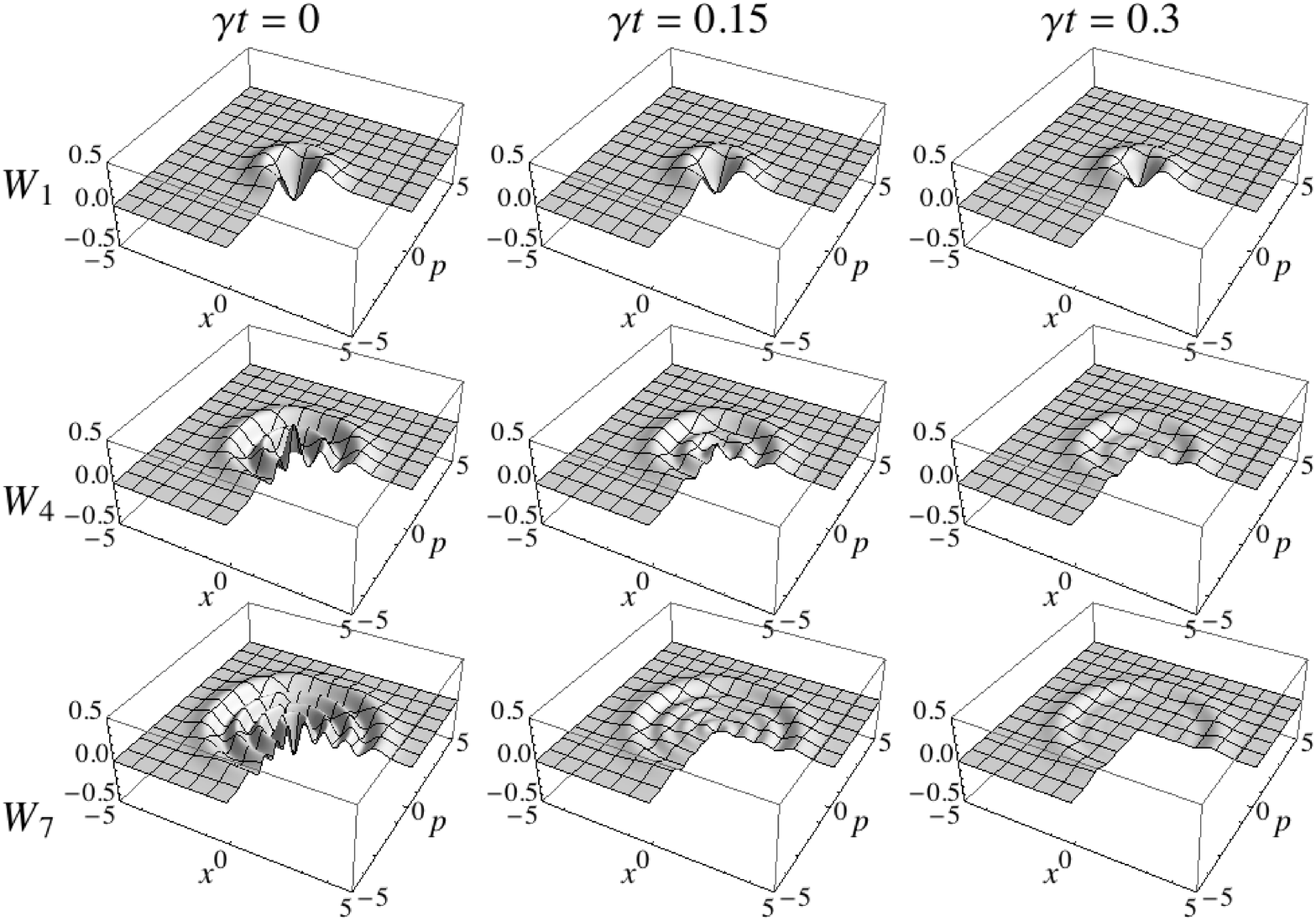}
    \label{fig:9pl}
   }
   \end{flushleft}
   \end{minipage}
   \vspace{0.5cm}
   \begin{minipage}[b]{0.3\textwidth}
    \begin{flushright}
    \subfigure[]{
    \includegraphics[width=\textwidth]{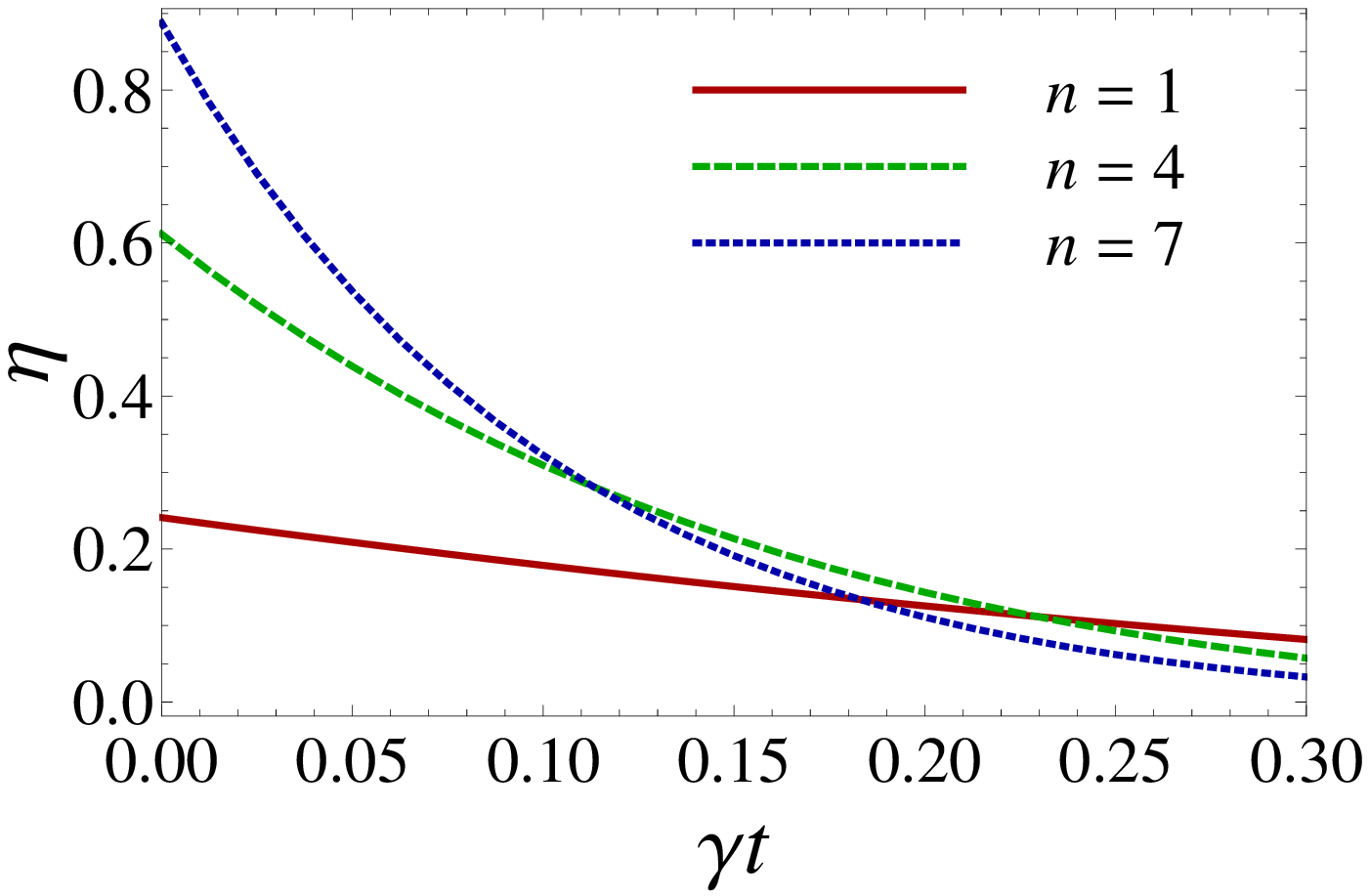}
    \label{fig:nvol}
    }\\
    \subfigure[]{
    \includegraphics[width=\textwidth]{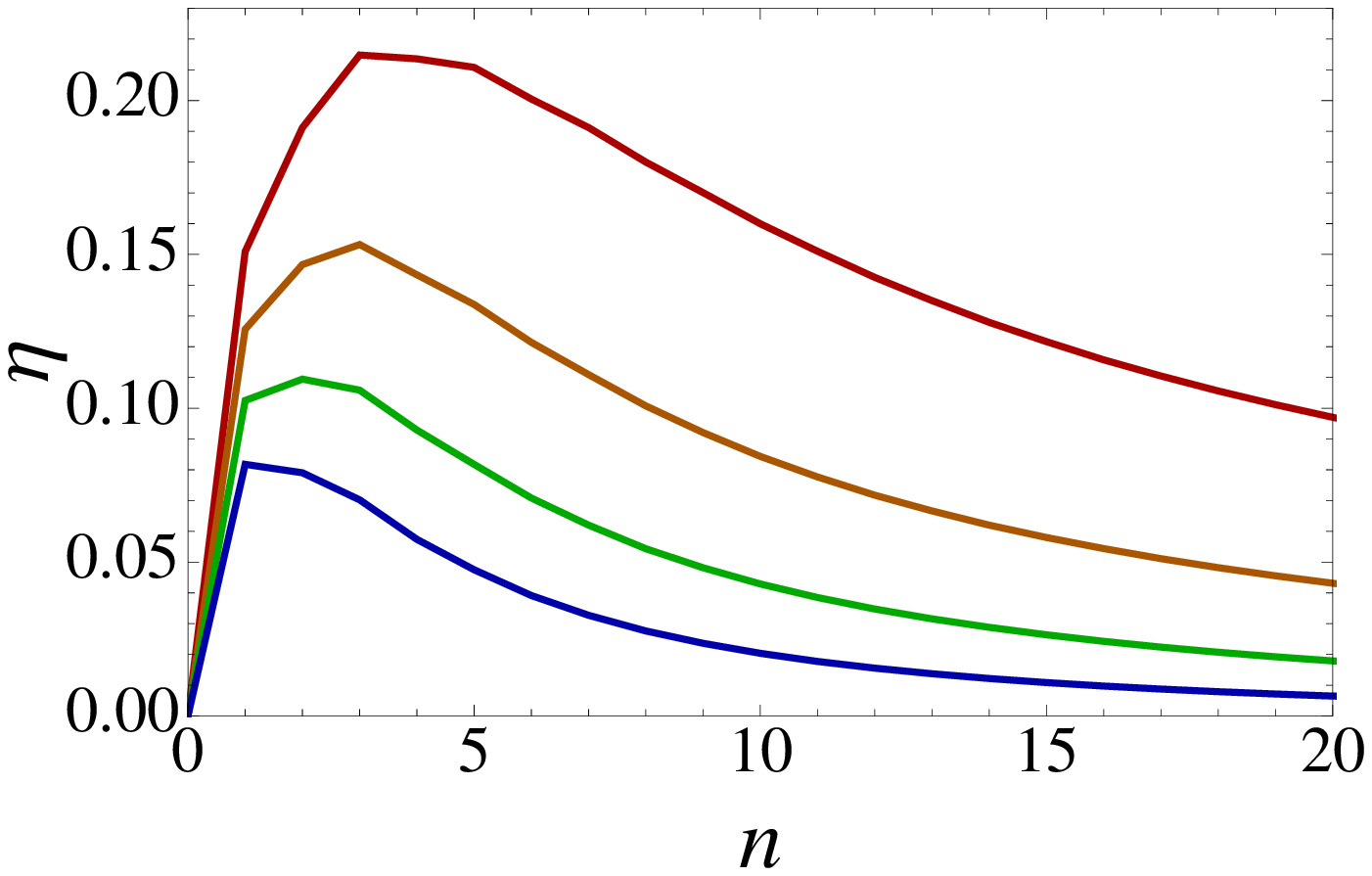}
    \label{fig:nvol2}
    }
   \end{flushright}
   \end{minipage}
   \caption{
    \textbf{(a)} Evolution of the Wigner function representations of a harmonic oscillator initialized in 
    the $n=1, 4$, and 7 Fock states. At $\gamma t=0$, the strongest quantum features are exhibited 
    by the $n=7$ state, but interaction with the environment yields a nearly classical distribution by 
    $\gamma t = 0.3$. The oscillator initialized in the $n=1$ state looks ``less quantum'' at 
    $\gamma t=0$, but retains its quantum features throughout the evolution, resulting in a 
    comparatively ``more quantum'' distribution at $\gamma t = 0.3$. (The fourth quadrant of the 
    $xy$-plane has been removed in order to emphasize the negative values obtained by the 
    Wigner function.) \textbf{(b)} (color online.) Negative volumes of the Wigner function 
    representations for oscillators initialized in the $n=1,4$ and 7 Fock states as a quantumness 
    metric to quantify the behavior observed qualitatively in (a). \textbf{(c)} (color online.)
    Comparison of the numerically-calculated negative volumes for harmonic oscillators initialized in 
    the 0--20th Fock states for four increasing values of $\gamma t$ showing a 
    time-dependent negative-volume peak. 
    The curves from top to bottom correspond to $\gamma t = .15$, $.20$, $.25$, and $.30$, 
    respectively, and the dashed lines indicate the systems 
    initialized in the $n=1,4$ and 7 Fock states.
   }
  \end{figure*}  
  
  Fig. \ref{fig:9pl} shows plots of the Wigner functions for zero-temperature harmonic 
  oscillators initialized in the $n=1,4$ and 7 Fock states for three increasing values of $\gamma t$. 
  (Note that allowing the system to evolve for a longer time $t$ at a given degree of coupling $
  \gamma$ is equivalent to allowing the system to evolve for a shorter time at a higher degree of 
  coupling---increasing $\gamma t$ captures either of these behaviors.) When $\gamma t=0$, the 
  system initialized in the $n=7$ Fock state exhibits strongly quantum features: its Wigner function 
  has large-amplitude oscillations and contains large regions of negative quasi-probability, an 
  obvious indicator of non-classical behavior. In these respects, the system initialized in the $n=4$ 
  Fock state is comparatively ``less quantum'' at $\gamma t=0$, and the system initialized in the 
  $n=1$ Fock state even less so.
  
  At $\gamma t = 0.15$, the oscillation amplitudes of the system initialized in the $n=7$ Fock state 
  are greatly reduced due to environmental interaction. The influence of the environment is present 
  in the other two systems as well, but the effect is less pronounced. By $\gamma t=0.3$, the 
  systems initialized in the $n=4$ and 7 Fock states resemble classical orbits.  Though it began as 
  the ``least quantum'' state, by $\gamma t = 0.3$, the Wigner function of the system initialized in the 
  $n=1$ Fock  state has retained the comparatively ``most quantum'' oscillations and negative 
  quasi-probability.
  
  A numerical measure of quantumness for unipartite systems is necessary in this context to 
  compare behavior.  There are various ways of parameterizing nonclassicality---it is 
  typical to define a coherent state with minimum uncertainty as the most classical system and to 
  consider the quantumness of a state in question to be the minimum distance to the coherent state, 
  using a distance metric based on the trace, the Hilbert-Schmidt distance, or a similar metric 
  \cite{dodonov00}. 
  It is also possible to characterize quantum states based on the properties of their Wigner function
  representations, an attractive option in light of recent successes reconstructing the Wigner 
  function experimentally \cite{lvovsky01, deleglise08}. Naively, one might consider the purity, 
  $\mathcal{P}$, defined in phase space as
   \begin{equation}
    \mathcal{P} = 2\pi \iint\limits_{\mathbb{R}^2} dxdp\left[W(x,p)\right]^2,
   \end{equation}
  to be a good indicator of quantumness. However, at zero temperature the time-asymptotic 
  distribution for any initial Fock state is $|0\rangle$. The purity of $|0\rangle$ is unity, but its Wigner 
  representation $\mathcal{W}_0(x,p)$ is gaussian, resembling a very classical distribution.
  
  Kenfack and $\rm{\dot{Z}}$yczkowski \cite{kenfack04} 
  suggest that the nonclassicality, or quantumness of a system can be captured using the 
  negative volume of the Wigner function, here denoted $\eta$ and given by
   \begin{equation}
    \eta(n,t) = \iint\limits_{\mathbb{R}^2}dxdp\frac{|W_n(x,p,t)|-W_n(x,p,t)}{2}.
    \label{eq:eta}
   \end{equation}
  Eq. \eqref{eq:eta} can be numerically integrated to obtain the negative volume of the 
  zero-temperature Wigner functions. In Fig. \ref{fig:nvol} the negative volumes for the oscillators 
  initialized in the $n=1,4$ and 7 Fock states are plotted as a function of $\gamma t$.

  \begin{figure}[t]
  \subfigure[]{
    \includegraphics[width=.47\textwidth]{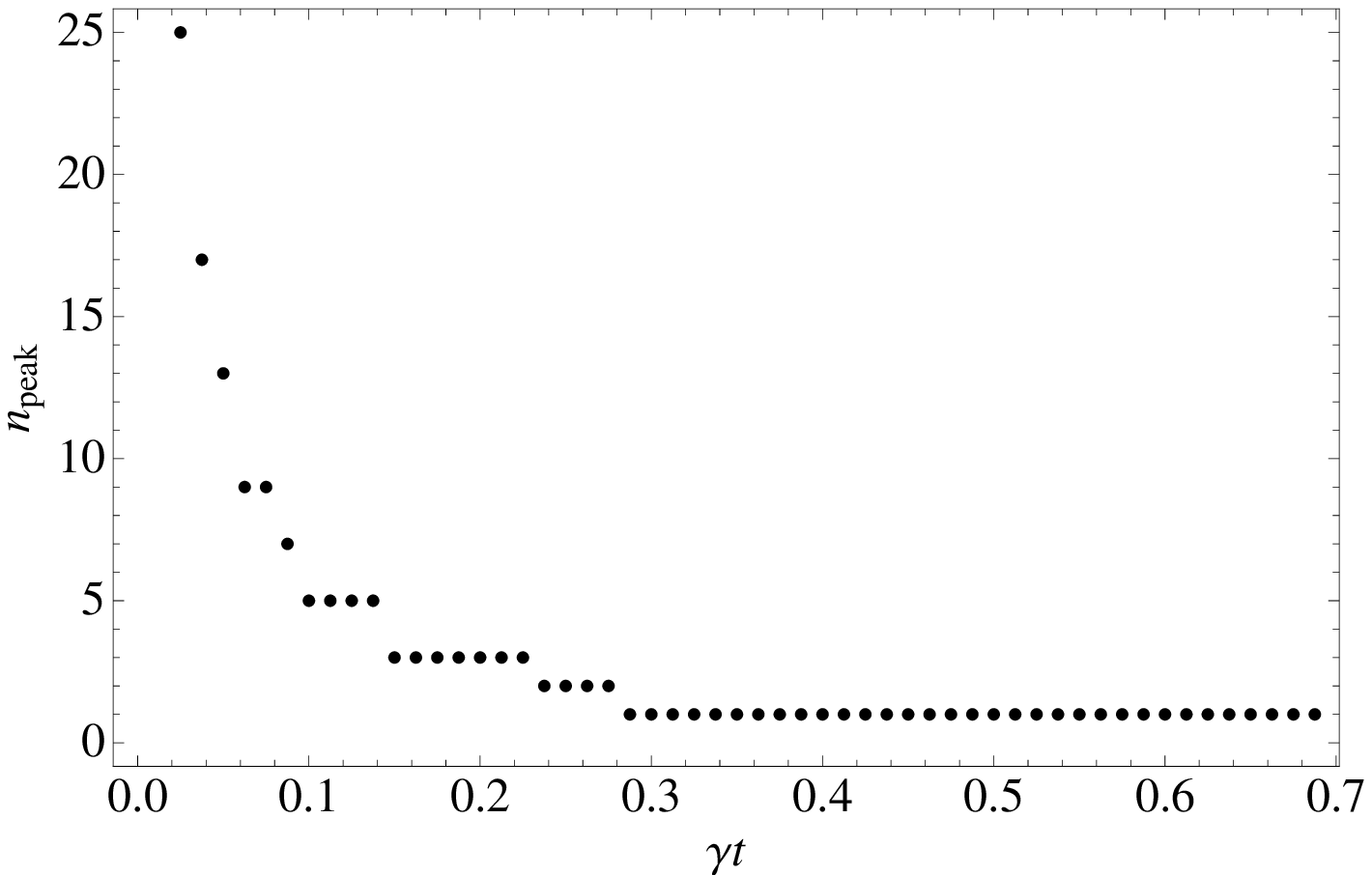}
    \label{fig:negt}   
  }
  \subfigure[]{
    \includegraphics[width = .47\textwidth]{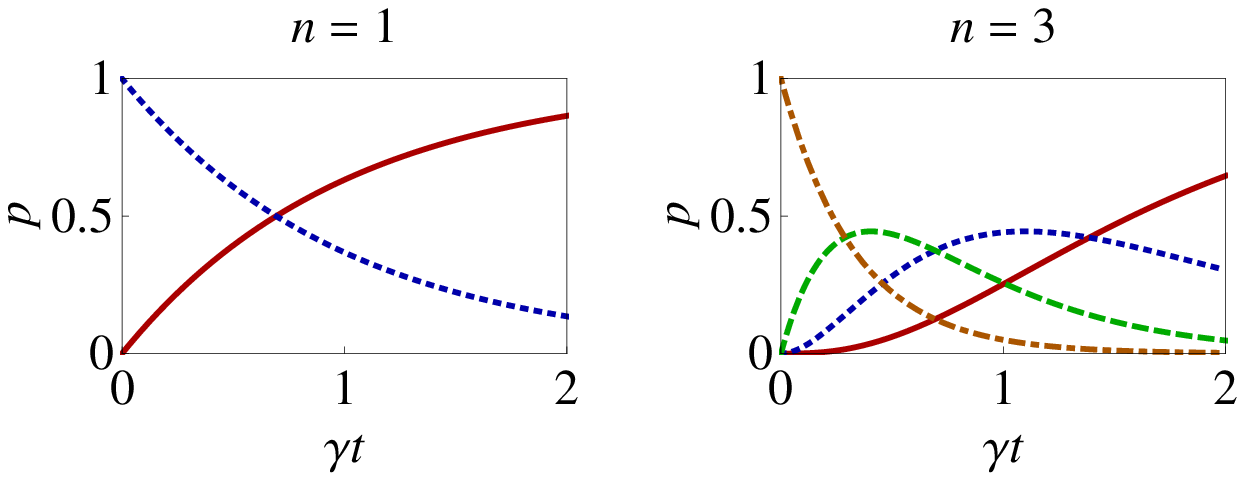}
   \label{fig:p}
   }
   \caption{
    \textbf{(a)} Peak quantum eigenstate as a function of $\gamma t$.
    \textbf{(b)} (color online.) Time-dependent occupation probabilities for oscillators initialized in 
    the $n=1$ ({\it left}) and $n=3$ ({\it right}) Fock states. The solid, dashed, dotted, and dot-dashed 
    lines are the occupation probabilities for the $n=0,1,2$ and $3$ states, respectively. These 
    curves strongly resemble those that have been determined experimentally 
    \cite{brune08,wang08}.
   }	
  \end{figure}

  Fig. \ref{fig:nvol2} shows the negative volume for the 0--20th Fock states at four different values 
  of $\gamma t$. For $\gamma t=0$, $\eta$ increases monotonically across the Fock states 
  (approximately as $\sqrt{n}$), as reported by Kenfack and $\rm{\dot{Z}}$yczkowski 
  \cite{kenfack04}. As the system evolves, $\eta$ obtains a finite-$n$ peak value, because though 
  oscillators initialized in higher Fock states are ``more quantum'' at the outset, the rate of 
  decoherence for larger $n$ states is higher. Fig. \ref{fig:negt} shows the peak quantum 
  state as a function of $\gamma t$. The existence of a quantumness peak can be understood
  precisely as resulting from these two competing trends: (a) the tendency for the fringing 
  behavior (and therefore the negative volume) to increase with initial $n$, and (b) the robustness 
  against decoherence to decrease with $n$. This behavior is arguably independent both of the 
  measure of quantumness and of the system under consideration.
  
  As mentioned above, the Wigner function for the interacting oscillator is composed exclusively of 
  diagonal elements in the harmonic oscillator basis. Its density matrix representation can therefore 
  be written $\rho_n = \sum_{k=0}^np_k(t)|k\rangle\langle k|$, where $p_k(t)$ is the probability of 
  observing the system in the $k$th eigenstate at time $t$. Since the density matrix is composed 
  of mutually orthogonal states, we can use the phase space trace  
  to determine the probabilities $p_k(t)$ \cite{schleich01}. For an oscillator initialized in the $n$th 
  Fock state
   \begin{equation}
    p_k(t) = e^{-n\gamma t}{n\choose k}(e^{\gamma t}-1)^{n-k},\quad k\leq n.
   \end{equation}
   
  Because the density matrix contains no off-diagonal elements, these probabilities are a complete 
  description of the quantum state. The behavior of $p_k(t)$ for oscillators initialized in the $n=1$ 
  and $n=3$ Fock states are shown in Fig.\ \ref{fig:p}. The occupation probabilities for 
  decohering Fock states up to $n=15$ have been determined experimentally \cite{brune08, 
  wang08}. The strong resemblance of the curves in Fig.\ \ref{fig:p} to experimental curves 
  \footnote{Fig. 3, Ref. \cite{brune08}; Fig. 4, Ref. \cite{wang08}.} suggests that evidence for a 
  quantumness peak has already been observed.
    
 In conclusion, the Wigner function representation of decohering zero-temperature Fock states can 
 be used to compare the quantumness of systems, using the negative volume of the Wigner 
 function as a quantumness metric. Though systems initialized in higher-$n$ Fock states start out 
 as the ``more quantum'' states, they decohere more quickly than lower-$n$ states; the competing 
 trends of quantum number and decoherence give rise to a time-dependent quantumness peak. 
 The general nature of this argument suggests that this quantumness peak may be a universal 
 feature of unipartite systems.
 
We thank Eric Egge and Bill McClung for useful discussions. Funding for this research from the 
Howard Hughes Medical Institute and the Clinton Ford Research Fund is gratefully acknowledged.
 
 \bibliography{bib2}

\begin{thebibliography}{15}
\expandafter\ifx\csname natexlab\endcsname\relax\def\natexlab#1{#1}\fi
\expandafter\ifx\csname bibnamefont\endcsname\relax
  \def\bibnamefont#1{#1}\fi
\expandafter\ifx\csname bibfnamefont\endcsname\relax
  \def\bibfnamefont#1{#1}\fi
\expandafter\ifx\csname citenamefont\endcsname\relax
  \def\citenamefont#1{#1}\fi
\expandafter\ifx\csname url\endcsname\relax
  \def\url#1{\texttt{#1}}\fi
\expandafter\ifx\csname urlprefix\endcsname\relax\def\urlprefix{URL }\fi
\providecommand{\bibinfo}[2]{#2}
\providecommand{\eprint}[2][]{\url{#2}}

\bibitem[{\citenamefont{Kenfack and
  $\mathrm{\dot{Z}}$yczkowski}(2004)}]{kenfack04}
\bibinfo{author}{\bibfnamefont{A.}~\bibnamefont{Kenfack}} \bibnamefont{and}
  \bibinfo{author}{\bibfnamefont{K.}~\bibnamefont{$\mathrm{\dot{Z}}$yczkowski}%
}, \bibinfo{journal}{J. Opt. B} \textbf{\bibinfo{volume}{6}},
  \bibinfo{pages}{396} (\bibinfo{year}{2004}).

\bibitem[{\citenamefont{Zurek}(2003)}]{zurek03}
\bibinfo{author}{\bibfnamefont{W.~H.} \bibnamefont{Zurek}},
  \bibinfo{journal}{Rev. Mod. Phys.} \textbf{\bibinfo{volume}{75}},
  \bibinfo{pages}{715} (\bibinfo{year}{2003}).

\bibitem[{\citenamefont{Blencowe}(2004)}]{blencowe04}
\bibinfo{author}{\bibfnamefont{M.}~\bibnamefont{Blencowe}},
  \bibinfo{journal}{Science} \textbf{\bibinfo{volume}{304}},
  \bibinfo{pages}{56} (\bibinfo{year}{2004}).

\bibitem[{\citenamefont{{F. Brennecke {\it et al.}}}(2008)}]{brennecke08}
\bibinfo{author}{\bibnamefont{{F. Brennecke {\it et al.}}}},
  \bibinfo{journal}{Science} \textbf{\bibinfo{volume}{322}},
  \bibinfo{pages}{235} (\bibinfo{year}{2008}).

\bibitem[{\citenamefont{van~der Wal~{\it et al.}}(2000)}]{wal00}
\bibinfo{author}{\bibfnamefont{C.~H.} \bibnamefont{van~der Wal~{\it et al.}}},
  \bibinfo{journal}{Science} \textbf{\bibinfo{volume}{290}},
  \bibinfo{pages}{773} (\bibinfo{year}{2000}).

\bibitem[{\citenamefont{Mabuchi and Doherty}(2002)}]{mabuchi02}
\bibinfo{author}{\bibfnamefont{H.}~\bibnamefont{Mabuchi}} \bibnamefont{and}
  \bibinfo{author}{\bibfnamefont{A.~C.} \bibnamefont{Doherty}},
  \bibinfo{journal}{Science} \textbf{\bibinfo{volume}{298}},
  \bibinfo{pages}{1372} (\bibinfo{year}{2002}).

\bibitem[{\citenamefont{{M. Brune {\it et al.}}}(2008)}]{brune08}
\bibinfo{author}{\bibnamefont{{M. Brune {\it et al.}}}},
  \bibinfo{journal}{Phys. Rev. Lett.} \textbf{\bibinfo{volume}{101}},
  \bibinfo{pages}{240402} (\bibinfo{year}{2008}).

\bibitem[{\citenamefont{{H. Wang {\it et al.}}}(2008)}]{wang08}
\bibinfo{author}{\bibnamefont{{H. Wang {\it et al.}}}}, \bibinfo{journal}{Phys.
  Rev. Lett.} \textbf{\bibinfo{volume}{101}}, \bibinfo{pages}{240401}
  (\bibinfo{year}{2008}).

\bibitem[{\citenamefont{Gardiner and Zoller}(2000)}]{gardiner00}
\bibinfo{author}{\bibfnamefont{C.~W.} \bibnamefont{Gardiner}} \bibnamefont{and}
  \bibinfo{author}{\bibfnamefont{P.}~\bibnamefont{Zoller}},
  \emph{\bibinfo{title}{Quantum Noise}} (\bibinfo{publisher}{Springer},
  \bibinfo{year}{2000}).

\bibitem[{\citenamefont{Serafini et~al.}(2004)\citenamefont{Serafini, Siena,
  and Illuminati}}]{serafini04}
\bibinfo{author}{\bibfnamefont{A.}~\bibnamefont{Serafini}},
  \bibinfo{author}{\bibfnamefont{S.~D.} \bibnamefont{Siena}}, \bibnamefont{and}
  \bibinfo{author}{\bibfnamefont{F.}~\bibnamefont{Illuminati}},
  \bibinfo{journal}{Mod. Phys. Lett. B} \textbf{\bibinfo{volume}{18}},
  \bibinfo{pages}{687} (\bibinfo{year}{2004}).

\bibitem[{\citenamefont{Benjamin and Quinn}(2003)}]{benjamin03}
\bibinfo{author}{\bibfnamefont{A.~T.} \bibnamefont{Benjamin}} \bibnamefont{and}
  \bibinfo{author}{\bibfnamefont{J.~J.} \bibnamefont{Quinn}},
  \emph{\bibinfo{title}{Proofs that Really Count: The Art of Combinatorial
  Proof}} (\bibinfo{publisher}{MAA}, \bibinfo{year}{2003}).

\bibitem[{\citenamefont{{V. V. Dodonov {\it et al.}}}(2000)}]{dodonov00}
\bibinfo{author}{\bibnamefont{{V. V. Dodonov {\it et al.}}}},
  \bibinfo{journal}{J. Mod. Opt.} \textbf{\bibinfo{volume}{47}},
  \bibinfo{pages}{633} (\bibinfo{year}{2000}).

\bibitem[{\citenamefont{{A. I. Lvovsky {\it et al.}}}(2001)}]{lvovsky01}
\bibinfo{author}{\bibnamefont{{A. I. Lvovsky {\it et al.}}}},
  \bibinfo{journal}{Phys. Rev. Lett.} \textbf{\bibinfo{volume}{87}},
  \bibinfo{pages}{050402} (\bibinfo{year}{2001}).

\bibitem[{\citenamefont{{S. Del\'eglise {\it et al.}}}(2008)}]{deleglise08}
\bibinfo{author}{\bibnamefont{{S. Del\'eglise {\it et al.}}}},
  \bibinfo{journal}{Nature} \textbf{\bibinfo{volume}{455}},
  \bibinfo{pages}{510} (\bibinfo{year}{2008}).

\bibitem[{\citenamefont{Schleich}(2001)}]{schleich01}
\bibinfo{author}{\bibfnamefont{W.~P.} \bibnamefont{Schleich}},
  \emph{\bibinfo{title}{Quantum Optics in Phase Space}}
  (\bibinfo{publisher}{Wiley-VCH}, \bibinfo{address}{Berlin},
  \bibinfo{year}{2001}).

\end{thebibliography}
  
\end{document}